\documentclass[namedreferences]{solarphysics}
\usepackage[optionalrh]{spr-sola-addons} % For Solar Physics
\usepackage{epsfig}          % For eps figures, old commands
\usepackage{graphicx}        % For eps figures, newer & more powerfull
\usepackage{color}           % For color text: \color command
\usepackage{url}             % For breaking URLs easily trough lines
\newcommand{\aap}{    {\it Astron. Astrophys.}}

\newcommand{\apj}{    {\it Astrophys. J.}}

\newcommand{\mnras}{  {\it Mon. Not. Roy. Astron. Soc.}}

\newcommand{\solphys}{{\it Solar Phys.}}
\newcommand{\sovast}{ {\it Soviet  Astron.}}
\newcommand{\ssr}{    {\it Space Sci. Rev.}}

\begin{document}

\begin{article}

\begin{opening}

%\title{Observation of an unusual solar radio burst at decameter wavelengths}

\title{Unusual Solar Radio Burst Observed at Decameter Wavelengths}

\author{V.N.~\surname{Melnik}$^{1}$\sep
        A.I.~\surname{Brazhenko}$^{2}$\sep
        A.A.~\surname{Konovalenko}$^{1}$\sep
        H.O.~\surname{Rucker}$^{3}$\sep
        A.V.~\surname{Frantsuzenko}$^{2}$\sep
        V.V.~\surname{Dorovskyy}$^{1}$\sep
        M.~\surname{Panchenko}$^{3}$\sep
        A.A.~\surname{ Stanislavskyy}$^{1}$
       }
\runningauthor{Melnik et al.}
\runningtitle{Unusual solar radio burst}

   \institute{$^{1}$ Institute of Radio Astronomy, National Academy of
                     Sciences of Ukraine, Kharkov, Ukraine
                     email: \url{melnik@ri.kharkov.ua}\\
              $^{2}$ Gravimetrical Observatory, National Academy of
                     Sciences of Ukraine, Poltava, Ukraine
                     email: \url{brazhai@gmail.com} \\
              $^{3}$ Space Research Institute, Austrian Academy of Sciences,
                     Graz, Austria
                     email: \url{mykhaylo.panchenko@oeaw.ac.at}
                     email: \url{rucker@oeaw.ac.at} \\
             }

\begin{abstract}
An unusual solar burst was observed simultaneously by two decameter
radio telescopes UTR-2 (Kharkov, Ukraine) and URAN-2 (Poltava,
Ukraine) on 3 June 2011 in the frequency range 16-28 MHz. The
observed radio burst has some unusual properties, which are
not typical for the other types of solar radio bursts. The
frequency drift rate of it was positive (about 500 kHz s$^{-1}$)
at frequencies higher than 22 MHz and negative (100 kHz s$^{-1}$) at
lower frequencies. The full duration of this event varies
from 50 s up to 80 s, depending on the frequency. The maximum radio flux
of the unusual burst reaches $\approx 10^3$ s.f.u and its
polarization does not exceed 10\%. This burst has a fine
frequency-time structure of unusual appearance. It consists of
stripes with the frequency bandwidth 300-400 kHz. We consider that several
accompanied radio and optical events observed by SOHO and STEREO
spacecraft are possibly associated with the reported radio burst.
A model that may interpret the observed unusual solar radio burst
is proposed.

%%\verb+SOLA_keyword_list.txt+.
\end{abstract}

\end{opening}

%-------------------------------------------------

\section{Introduction}
     \label{Introduction}

Sporadic radio emission of the Sun is observed all over the wavelength
band, from millimeter to kilometer wavelengths \cite{Suzuki1985}. The observed frequencies of some
types of radio bursts such as type II and type III bursts as well as S-bursts drift from high to low frequencies,
whereas other radio emissions, $\:i.e.$ type IV radio bursts, are recorded simultaneously in a wide frequency
range from decimeter to decameter wavelengths. On the other hand narrow and short spikes of emission
are observed independently at decimeter, meter, and decameter wavelengths.

In the decameter range the following types of sporadic solar radio emission are observed

 \begin{itemize}
   \item [-]type III bursts \cite{Abranin1980,Melnik2005a} and, related to them, type IIIb bursts \cite{Bazelian1978,Melnik2010a};
   %\item [-] fast type III bursts (Melnik et al., 2008a);
   %\item [-] "dog-leg" type III bursts (Dorovskyy et al., 2011);
   \item [-] ordinary type II bursts and  type II bursts with herringbone
             structure \cite{Melnik2004};
   \item [-] type IV bursts \cite{Melnik2008b};
   \item [-] inverted U- and J- bursts \cite{Dorovskyy2010};
   \item [-] S-bursts (Dorovskyy \emph{at al.}, 2006; Briand \emph{at al.}, 2008; Melnik \emph{at al.}, 2010b)\nocite{Dorovskyy2006,Briand2008,Melnik2010b};
   \item [-] drift pairs \cite{Melnik2005b};
   \item [-] short and extended bursts in absorption \cite{Konovalenko2007,Melnik2010c}.
 \end{itemize}

 %1) type III bursts (Abranin et al., 1980, Melnik et al., 2005a) and related
%    to them type IIIb bursts (Bazelian et al., 1978, Melnik et al., 2010a);
% 2) fast type III bursts (Melnik et al., 2008a);
% 3) "dog-leg" type III bursts (Dorovskyy et al., 2011);
% 4) usual type II bursts and  type II bursts with herring born structures
%    (Melnik et al., 2004);
% 5) type IV bursts (Melnik et al., 2008b);
% 6) inverted U- and J- bursts (Dorovskyy et al., 2010);
% 7) S-bursts (Dorovskyy et al., 2006, Briand et al., 2008,
%    Melnik et al., 2010b), drift pairs (Melnik et al., 2005b),
% 8) short and extended bursts in absorption (Melnik et al., 2010c, Konovalenko et al., 2007)

The decameter radio telescopes with large effective areas
such as UTR-2 equipped with spectrometers operating in a broad
frequency band have discovered fast type III bursts
\cite{Melnik2008a}, dog-leg type III bursts \cite{Dorovskyy2011},
and decameter spikes \cite{Melnik2011}. Moreover the modern
back-end facilities with high time-frequency resolution allowed us
to study the properties of fine temporal structure of type
II and III bursts (Melnik {\it et al.}, 2004, 2005a\nocite{Melnik2004}\nocite{Melnik2005b}) and zebra structure of type IV
bursts \cite{Melnik2008b} in details.

The main parameters of the solar radio bursts are the frequency band width,
frequency drift rate, duration, polarization degree, and
flux of the radio emission. This set of parameters allows us to
classify radio bursts. For example, a typical
frequency drift rate of the usual (not fast) type III bursts  in
the decameter range is 2-4 MHz s$^{-1}$, whereas type II bursts have
frequency drift rates about 30-70 kHz s$^{-1}$ \cite{Melnik2004}. The duration
of type III bursts in the decameter band is 6-12 s
\cite{Abranin1980, Melnik2005a}. Type IV bursts are observed
during an interval from 1-1.5 h to several hours
\cite{Melnik2008b}, whereas S-bursts have duration of only
0.3-0.6 s. The frequency drift rate of S-bursts is about 1 MHz s$^{-1}$
\cite{Melnik2010b}.

The sources of radio emission for different types of bursts are fast particles (in most cases,
electrons), which propagate through different magnetic structures in the solar corona
\cite{Suzuki1985}. Fast electrons accelerated during flares and moving along open
the magnetic field lines are responsible for type III bursts. Electrons accelerated at shock
fronts generate type II radio bursts. Radio emission of type IV bursts is a result of
interaction between CME and plasma of the solar corona.

In this paper we report the observation of an unusual solar radio burst. This burst was
recorded simultaneously by two Ukrainian decameter radio telescopes, UTR-2
(Kharkov, Ukraine) and URAN-2 (Poltava, Ukraine) on 3 June 2011.
The observed solar radio burst cannot be referred to any known type of the
bursts in the decameter  range. We discuss the properties of this
burst and consider the associated radio and plasma phenomena observed simultaneously
by SOHO and STEREO. A model or possible scenario of the unusual burst generation is proposed.

\section{Observations of Unusual Decameter Solar Radio Burst} %%%%%%%%%%%%%%%%%%%%%%%%%%%%%%%%%%%%%%%%
      \label{telescopes}

An unusual burst in the form of a ``caterpillar" was observed during
a decameter type III storm at 12:10 on 3 June  2011. This burst was
observed simultaneously by UTR-2 and URAN-2 radio telescopes.
During this observation the UTR-2 radio telescope was operated in
the mode in which only four sections of the north-south branch of the
antenna was used. The total effective area of these four
sections was 50000 m$^2$ with the beam pattern size of $1^{\circ}
\times 15^{\circ}$ at 25 MHz. The radio telescope URAN-2 (Poltava, Ukraine)
has the effective area of 28000 m$^2$ and its beam pattern size
is $3.5^{\circ} \times 7^{\circ}$ at 25 MHz \cite{Megn2003,Brazhenko2005}.
Additionally URAN-2 is able to  measure the degree of polarization of the
received radio signal. The data were recorded by the digital
DSP spectrometers \cite{Ryabov2010} operating in the
frequency range of 16-32 MHz with a frequency-time resolution of $4 \:
\mathrm{kHz} \times 100 \: \mathrm{ms}$.

Figure 1 shows  the dynamic radio spectra obtained by URAN-2
(panel a) and UTR-2 (panel b). The unusual burst was observed at
around 12:10 UT at the frequencies from $\approx 16 \:\mathrm{MHz}$
to $\approx 27.5 \:\mathrm{MHz}$. This burst was recorded
simultaneously by UTR-2 and URAN-2 radio telescopes and the obtained
radio spectra look identically.

High frequency-time resolution of the radio telescopes
gives an opportunity to observe a complex fine structure of the
unusual burst as shown in Figure 2. The burst consisted of 14
narrow-band stripes of different duration. Comparing this stripe
structure with known striae fine features of type IIIb
bursts we can find significant differences. In particular, the
frequency bandwidth of the stripes of the unusual burst is about
300-400 kHz whereas the striae of type IIIb bursts in the
decameter range have narrower bandwidth, only 50-70 kHz
\cite{Melnik2011}. The second difference is that individual
stria of type IIIb present a straight track whereas the stripes of
unusual burst looked like a serpentine.

The caterpillar exhibited negative and positive frequency
drifts simultaneously. In particular, the frequency drift rate of the burst was
positive (about 500 kHz s$^{-1}$) at frequencies higher than 22 MHz and
negative (100 kHz s$^{-1}$) at lower frequencies. These frequency drift
rates are lower than those for type III bursts but higher
than typical drift rates for type II bursts in the decameter
range. The maximum flux of the unusual radio burst did not exceed
$10^3$ s.f.u.

The very interesting feature of the caterpillar is its strong
cut-off at high frequencies. As shown in Figure 1 the burst
abruptly disappeared at 27.5 MHz. At the same time there was no
cut-off at low frequencies and the burst was observed by STEREO
significantly below 16 MHz (see Figure 5).

The time profile of the unusual burst at 26 MHz is shown in Figure 3.
Contrary to type III bursts the time profile of the unusual burst
had a bell-like shape. The full duration of the burst varied from 50 s  at lower
frequencies to 80 s at frequencies above 22 MHz. Such durations are significantly
larger than the average durations of individual type III bursts at decameter wavelengths, namely  6-12 s.

Figure 4 shows the time profile of the polarization measurements at
a frequency of 26 MHz. The polarization degree of the caterpillar
was practically constant and was $\approx 10\%$. Such a degree of
polarization is typical for the second harmonic of type III
bursts \cite{Dulk1980}. This give us an idea that if the
generation mechanism of the unusual burst is similar to those for
type III bursts, then the unusual burst was radiated at the second
harmonic.

Additional to the ground-based observations we have also examined
the data recorded by radio experiment WAVES onboard two STEREO
spacecraft (STEREO-A and STEREO-B; Bougeret \emph{at al.}, 2008\nocite{Bougeret2008}). On 3 June
2011 two spacecraft were located on the heliocentric orbits
with angular separation $\approx 180^{\circ}$, \emph{i.e.} on the opposite sides
of the Sun (see Figure 5b). The dynamic radio spectra recorded by
STEREO-A  and STEREO-B in the frequency range from 1 MHz to 16 MHz
are shown in Figure 5a. As is seen, only one
spacecraft, STEREO-A, observed the unusual burst at $\approx $
12:10 UT. STEREO-B recorded only a group of type III bursts at
frequencies lower than 2 MHz at around 12:04 UT, which
was also well visible in the UTR-2 and URAN-2 spectra (see Figure 1). Taking into
account the positions of STEREO-A  and STEREO-B (Figure 5b) we conclude that the
source of the unusual burst was situated on the west side of the Sun with respect to the Earth.
We have also estimated the frequency drift rate of the unusual burst observed by
STEREO-A at frequencies 1-16 MHz. This drift rate was about 50 kHz s$^{-1}$
which was smaller than typical drift rates of type III bursts at these frequencies (see, for
example, a group of type III bursts at 12:04 in Figure 5a).

\section{Observations of Accompanied Phenomena}

After detection of the caterpillar (at 12:10 UT) two
coronagraphs LASCO C2 and C3 onboard SOHO have observed three
phenomena on the west side of the solar disk (Figure 6).
In particular LASCO C2 recorded two successive
ejections appeared at 12:48 and 13:36 UT. The sizes of these two ejections were
smaller than the typical size of coronal mass
ejections (CMEs). At 12:48 there was also a jet at a latitude higher than
those of the ejections (Figure 6). According to UV images from SOHO
(Figure 7a) no active regions were detected on the west side of
the solar disk. At the same time STEREO-A/EUVI observed two active
regions NOAA 1224 and NOAA 1222 situated at approximately
$100^\circ$ and  $130^\circ - 140^\circ $ of western longitude
with respect to the central meridian (Figure 7b). Since NOAA 1222 was
on the lower latitude than the active region NOAA 1224 we may consider that
NOAA 1222 initiated two ejections while the NOAA 1224 triggered the
jet.

In order to clarify which of the ejections or the jet observed by
SOHO (Figure 6) were related to the unusual  burst, let us consider
how the ejections and jet  propagated. Figures 8 and 9 present
the motions of two ejections as well as the front of the jet
according to LASCO C2 and LASCO C3 observations. From  LASCO C2
observations the front $ r_{\mathrm{f, p}}$  and the center  $r_{\mathrm{c, p}}$  of
the first ejection moved in the plane of the sky as

  \begin{eqnarray}
     r_{\mathrm{f, p}}=2.95 \times 10^{-2} \cdot t +1.79  \label{Eq1} \\
     r_{\mathrm{c, p}}=2.91 \times 10^{-2} \cdot t +1.45 \label{Eq2}
  \end{eqnarray}
where $r$ is the radial distance in units of the solar radius $R_\mathrm{s}$ , and  $t$ is the time in
minutes counted from 12:10 UT. At the same time according to
LASCO C3 data this propagation can be described as:

 \begin{eqnarray}
     r_{\mathrm{f, p}}=2.43 \times 10^{-2} \cdot t +1.89  \label{Eq3} \\
     r_{\mathrm{c, p}}=2.55 \times 10^{-2} \cdot t +1.003. \label{Eq4}
  \end{eqnarray}
The above equations as well as the following (Equations
(\ref{Eq5}) -- (\ref{Eq13})) are valid in the plane of the sky.

The linear velocities of the front and the center of the first
ejection derived from the above equations are, respectively, 344
km s$^{-1}$ and 340 km s$^{-1}$ near the Sun (LASCO C2) and 284 km s$^{-1}$ and 298
km s$^{-1}$ far from the Sun (LASCO C3). It is seen that the first
ejection decelerated with the distance, particularly the front of
the ejection  $ r_\mathrm{f, p}$  decelerated faster than the ejection
center $ r_\mathrm{c, p}$. At 12:10, when the unusual burst started at
a frequency of 22 MHz, the front of the first ejection was at $1.8 R_\mathrm{s}$
and $1.9 R_\mathrm{s}$  from the center of the Sun in the plane of the sky
according to LASCO C2 and C3 respectively. Taking into
account that this ejection was initiated by active region NOAA 1222
with a longitudinal angle $\alpha =40^{\circ} - 50^{\circ}$ (
$\alpha$  is defined in Figure 13), the real radial distance from
the front of the ejection to the center of the Sun was $2.25 - 2.55 R_\mathrm{s}$
(for $\alpha =40^{\circ}$) or $2.35 - 2.69 R_\mathrm{s}$ ( for $\alpha
=50^{\circ}$). At the same time the radio emission of the unusual
burst appeared firstly on the spectra at 22 MHz (at about 12:09:45
UT). Assuming that this radio emission was generated at the second
harmonic of the local plasma frequency and using the models of Newkirk (1961) and Baumbach-Allen
\cite{Allen1947} we can estimate that the radio source
should be at $r = 2.8 R_\mathrm{s}$   and  $r = 2.15 R_\mathrm{s}$, respectively, in these models.
Therefore we may suggest that the first ejection was very close
to the place from which the radio emission of the unusual burst came out.

Equations which define the motion of the front $r^{(2)}_\mathrm{f,p}$  and the center
$r^{(2)}_\mathrm{c,p}$ of the second ejection according to LASCO C2 observations are the following:

 \begin{eqnarray}
     r^{(2)}_\mathrm{f, p}=2.67 \times 10^{-2} \cdot t +0.19   \label{Eq5} \\
     r^{(2)}_\mathrm{c, p}=2.81 \times 10^{-2} \cdot t - 0.29 \label{Eq6}
  \end{eqnarray}
and according to LASCO C3 data:

  \begin{eqnarray}
     r^{(2)}_\mathrm{f, p}=2.37 \times 10^{-2} \cdot t +0.39   \label{Eq7} \\
     r^{(2)}_\mathrm{c, p}=2.61 \times 10^{-2} \cdot t - 0.66. \label{Eq8}
  \end{eqnarray}

Velocities (in the sky plane) of the front and the center of
the second ejection were 311 km s$^{-1}$ and 328 km s$^{-1}$ (according to LASCO
C2) and 277 km s$^{-1}$ and 305 km s$^{-1}$ (according to LASCO C3). For the
second ejection the velocities of the front and the center of the
ejection decreased with distance but only at the vicinity of the Sun,
whereas far from the Sun the center moved faster than the front.
As follows from Equations (\ref{Eq5}) -- (\ref{Eq8}) and Figure 8
the second ejection reached the estimated heights
in the corona ( $2.15 - 2.8 \:R_\mathrm{s}$) in 80-90 min after
the beginning of the unusual burst. Therefore we can conclude that
this ejection did not  cause the unusual radio burst.

The jet front propagated through the solar corona with a constant
velocity of about 265 km s$^{-1}$ as seen in Figure 9 (taking into account the longitudinal
angle $\alpha =10^{\circ}$ of the active region NOAA 1224 which was supposed
to associate with the jet). The equation which describes the motion  of the
jet front in the sky plane is:

 \begin{eqnarray}
     r^\mathrm{j}_\mathrm{f, p}=2.25 \times 10^{-2} \cdot t +2.27.   \label{Eq9}
  \end{eqnarray}

The distance from the front of the jet to the center of the Sun at the time of the unusual
radio burst was  $2.3 R_\mathrm{s}$. This radial distance
is close to those at which the radio emission at the second harmonic at 22 MHz can be generated.
Therefore the jet can be a source of the unusual radio burst too.
Nevertheless we suggest that the unusual burst was connected with
the first ejection because in this case some properties of the burst can be explained
easier.

Further we need to know the size of the first ejection at the
time of the unusual burst radiation. Figure 10 shows the dependence of
the longitudinal $(l)$  and the transverse $(d)$ sizes of the ejection
on the distance from the Sun according to LASCO C2 and C3 data.
These dependencies for LASCO C2 data can be formulated as follows:

  \begin{eqnarray}
     l_\mathrm{p} = 0.57 \times 10^{-2} \cdot t +0.37  \label{Eq10} \\
     d_\mathrm{p} = 0.15 \times 10^{-2} \cdot t +0.26  \label{Eq11}
  \end{eqnarray}
and out of LASCO C3 observations
  \begin{eqnarray}
     l_\mathrm{p} = 0.3 \times 10^{-2} \cdot t +0.63  \label{Eq12} \\
     d_\mathrm{p} = 0.21 \times 10^{-2} \cdot t +0.2.  \label{Eq13}
  \end{eqnarray}

The velocity estimated from the longitudinal extension of the ejection is $\upsilon_l$ = 66.5 km s$^{-1}$ which is four times larger
than the velocity derived from the transverse extension ($\upsilon_d$ = 17.5 km s$^{-1}$)
in the vicinity of the Sun, whereas far from the Sun
$\upsilon_l$ = 35 km s$^{-1}$ and $\upsilon_d$ = 24.5 km s$^{-1}$.
The expansion velocities of the first ejection were smaller than the
bulk velocity.  Therefore we can conclude that
the density of the ejecta practically did not change during its movement.
Such situation is possible when the surface magnetic field of the ejecta
restrains plasma from expanding into the surrounding coronal plasma.
Thus further we consider the ejection as a magnetic flux rope.

\section{Model of Unusual Burst}

We suppose that the magnetic flux rope was connected to the active region
NOAA 1222. As it follows from STEREO-A/EUVI observations this
region showed very high activity beginning from 11:36 UT. It manifested
in increasing its brightness  and size as well as in the appearance of
arches related to this region. These processes lasted for about 2 h.
Supposing that the magnetic flux rope moved radially from the active
region with the longitude angle $\alpha$ and using Equations
(\ref{Eq1}) and (\ref{Eq2}) we derive the following equations
which describe the motions of the front $r_\mathrm{f}$  and center $r_\mathrm{c}$
of the ejection

 \begin{eqnarray}
     r_\mathrm{f} = (2.95 \times 10^{-2} \cdot t +1.79) \cos^{-1} \alpha   \label{Eq14} \\
     r_\mathrm{c} = (2.91 \times 10^{-2} \cdot t +1.45) \cos^{-1} \alpha.  \label{Eq15}
 \end{eqnarray}
The magnetic flux rope was formed approximately at 11:36 UT
(at the very beginning of NOAA 1222 activity) at
longitudes $\alpha =40^{\circ} - 50^{\circ}$ (according to
Equation (\ref{Eq14}), $t=-34$ min) near the surface of the Sun.
The central part of the flux rope appeared in 6 min, \emph{i.e.} at 11:42 UT.
According to Equation (\ref{Eq10}) and taking into account the angle $\alpha
=40^{\circ} - 50^{\circ}$ the longitudinal size of the magnetic flux rope was
$l \approx 0.17 R_\mathrm{s}$ at the time of its formation. Velocities
of its front and center were, respectively, 450 km s$^{-1}$ and 443 km s$^{-1}$
at $\alpha =40^{\circ}$ and 535 km s$^{-1}$ and 528 km s$^{-1}$ at $\alpha
=50^{\circ}$. The ejection front was located at $R=1.8 R_\mathrm{s}/ \cos
\alpha = 2.22 - 2.55  R_\mathrm{s}$ from the center of the Sun at the
beginning of the unusual burst, \emph{i.e.} at 12:10 UT (see Figure 11). As was
discussed in Section 3, at these heights the local plasma
frequency is about 11 MHz. Therefore the magnetic flux rope at these
altitudes can be a source of radio emission at the second harmonic, 22 MHz.

Langmuir waves required for the generation of the radio emission
through the plasma mechanism can be excited by electrons which
are accelerated during the interaction of the moving magnetic flux rope
with the surrounding coronal plasma. In the case when the magnetic field of the flux rope
was perpendicular to the rope axis the electrons can be accelerated along the flux rope, for
example, by the Lorentz force ${\mathbf{F}_\mathrm{L}} = \frac{e}{c}
{\mathbf{v}_{{d}} \times {\mathbf{B}}_{r}} $, where
${\mathbf{v}_{d}}$ is the transverse velocity of the magnetic flux rope and
${\mathbf{B}_{r}}$  is the magnetic field of the flux rope (see
Figure 12a). The other possible acceleration process is magnetic
reconnection (Figure 12b) when the magnetic field of the flux rope was
oriented along the direction of the propagation of the flux rope. At the
constant acceleration along the flux rope the electrons can reach the
velocity  $\upsilon \approx 2L/ \Delta t$, where  $L$ is the
length of the flux rope and $\Delta t$  is the acceleration time.

Assuming that the acceleration time is approximately equal to the
duration of the unusual radio burst and taking into account that
the length of the flux rope is $\approx 0.5 R_\mathrm{s}$   at radial distances of $R=
2.22 - 2.55 R_\mathrm{s}$ we can estimate the velocities of fast electrons as
$\upsilon = 0.8 - 1.4 \cdot 10^9$ cm s$^{-1}$. These electrons
can propagate towards and away from the Sun. Electrons moving
through the coronal plasma generate Langmuir waves $l$  which in
the process $l+i = t+i$ ( $i$ means ion) are transformed into
electromagnetic waves $t$ at the first harmonic. In the
process  $l + l =t$ the Langmuir waves are merged into electromagnetic waves
at the second harmonic \cite{Ginzburg1958}. The radio emission at
the first harmonic generated in a behind-the-limb region cannot
propagate toward the Earth. At the same time the second harmonic
of the behind-the-limb radio emission can be observed on the Earth (see
the Appendix). In the Newkirk model of coronal density, the radio emission at the second
harmonic has a strong cut-off at the frequency $f^{\ast}$  (see
Equation (\ref{A4})). For the active regions with the longitudinal angle
$\alpha =42^{\circ}$ this frequency is $f^{\ast} =27.5 $ MHz. It
is the same frequency at which the unusual radio burst had a
clear cut-off (see Figure 1).

We propose the following explanation of this cut-off of the caterpillar
burst at 27.5 MHz. Electrons which were accelerated by
the magnetic flux rope at $R=2.22-2.55 R_\mathrm{s}$ towards the Sun generate the
radio emission of the unusual burst which drifted from  $22$
MHz up to $27.5$ MHz. The radio emission  could not propagate through
the solar corona in the direction to the Earth at frequencies  $f>
27.5$ MHz. At the same time electrons accelerated by the magnetic
flux rope away from the Sun produced the radio emission of the unusual burst
which drifted to low frequencies from 22 MHz to 16 MHz (according
data of UTR-2 and URAN-2) and down to 1 MHz (according to the STEREO A data).

\subsection{Frequency Drift Rate of the Unusual Burst}

As was mentioned in Section \ref{telescopes}, one of the interesting properties
of the caterpillar burst is its unusual frequency drift. In particular, the frequency drift
was positive at frequencies higher than 22 MHz and its rate was about 500 MHz s$^{-1}$,
whereas  at frequencies lower than 22 MHz the unusual burst exhibited negative
frequency drift with the rate of about 100 MHz s$^{-1}$.

The particles accelerated by the flux rope may produce the radio emission which will be observed
with the following frequency drift rate

 \begin{equation}\label{Eq16}
     \frac{df}{dt} = \frac{df}{dn} \cdot \frac{dn}{dr} \cdot \frac{\upsilon_\mathrm{s} c}{c-\upsilon_\mathrm{s}  \cos \alpha},
 \end{equation}
where $\upsilon_\mathrm{s}$  is the electron velocity, $c$ is the speed of light, and $n$  is the plasma density.
Using the Newkirk model we derive from Equation (\ref{Eq16}) in the case of the second harmonic of radio emission the following values for the electron velocities:
$\upsilon_\mathrm{s}=0.4 \cdot 10^9$ cm s$^{-1}$  and $\upsilon_\mathrm{s}=1.8 \cdot 10^9$ cm s$^{-1}$
for the observed drift rates  $\frac{df}{dt} =100$ kHz s$^{-1}$  and $\frac{df}{dt} =500$ kHz s$^{-1}$,
respectively.
These velocities are close to those obtained for the acceleration of electrons by the magnetic flux rope.

\subsection{Inhomogeneities of Coronal Plasma}

Our observations showed a fine structure of the unusual burst in the form of stripes (Figure 2).
These stripes had a longer duration than the striae of type IIIb bursts.
Their frequency bandwidths are 6-8 times broader than the bandwidths of striae, \emph{i.e.}
300-400 kHz for the unusual radio burst contrary to 50-70 kHz bandwidth of the type IIIb stria
\cite{Melnik2010a}. Additionally in contrast to type IIIb  bursts with a nearly constant frequency drift of
the striae, the drift rate of the unusual burst stripes changed during the stripe (Figure 2).

We believe that the discussed stripes were related to an inhomogeneity of coronal plasma.
This idea is analogous to the hypothesis proposed by \inlinecite{Takakura1975} according to which
the type IIIb stria radio emission may be explained as an amplification of the radio emission by
irregularities in the solar corona, for example, by the filaments.

\inlinecite{Kontar2001} has considered the propagation of the fast electron beams in a plasma
with density fluctuations and has shown that the Langmuir turbulence generated by electrons
is higher in the low plasma density regions.
Therefore, we can also expect more intense radio emission at regions of low density.
The size of inhomogeneties which are responsible for the enhancement of the radio stripes
of the unusual burst can be derived from the frequency bandwidth defined by

 \begin{equation}\label{Eq17}
     \Delta f \approx \frac{1}{\pi} \frac{d \omega_\mathrm{pe}}{dn} \frac{dn}{dr} \cdot \Delta r \approx \frac{1}{2 n} \frac{d n}{dr} f \cdot \Delta r.
 \end{equation}
From Equation (\ref{Eq17}) we obtain the following size of inhomogeneities  $\Delta r \approx  \Delta f \cdot R_\mathrm{s} /f \approx 10^{-2}R_\mathrm{s}$ .
Thus we can conclude that the electron beams propagate through essentially non-homogeneous plasma.
Note that an analogous result was derived for wide-band fibers, the fine structure elements of decameter type II bursts \cite{Chernov2007}.

\subsection{Brightness Temperature of the Unusual Radio Burst}

Proposing that the radio emission of the unusual burst was
generated by the particles accelerated by the magnetic flux rope
it is reasonable to accept that the transverse size $d$ of the volume of the
accelerated particles is equal to the transverse size of the
magnetic flux rope at the radial distances from which the radio
emission comes out. From Equation (\ref{Eq11}) we derive that
$d \approx 0.2 R_\mathrm{s}$. Taking into account that the maximum flux of
the unusual burst was about $F= 10^3$ s.f.u. (1 s.f.u. =
$10^{-22}$ W m$^{-2}$ Hz$^{-1}$ ) the corresponding brightness temperature is

 \begin{equation}\label{Eq18}
        T_\mathrm{b} = \frac{\lambda^2 F}{2 k \Omega}   \approx  10^{12} \: K
 \end{equation}
at the frequency $f=22$ MHz  and for the solid angle $\Omega = \pi d^2 /4 R_\mathrm{SE}^2=0.7 \cdot 10^{-6}$,
where $k$  is the Boltzman constant and $R_\mathrm{SE}$ is the distance from the Sun to the Earth.
Such a high brightness temperature can be explained in the frame of plasma mechanism of radio emission
\cite{Melnik2003}.

\section{Conclusions}

The unusual burst observed by radio telescopes UTR-2 and URAN-2 at frequencies 16-27 MHz
had the following properties:

\begin{enumerate}
  \item Radio emission came out from a behind-the-limb region.
  \item The burst has some unusual properties which are not typical for
         other known solar bursts. In particular,
           \begin{itemize}
            \item [-] the frequency drift rate was negative at low frequencies and positive at high frequencies;
            \item [-] the frequency drift rate changed in the range from 100 kHz s$^{-1}$ to 500 kHz s$^{-1}$;
            \item [-] the burst disappeared at frequencies $> 27.5$ MHz, \emph{i.e.} it showed strong frequency cut-off effect;
            \item [-] the total duration of the burst  (50 s - 80 s) was longer than that for the usual type III bursts;
            \item [-] the burst had a fine structure in the form of stripes with frequency bandwidth  $300-400$ kHz;
         \end{itemize}
  \item  The polarization of the burst was about 10\%, which corresponded to the second harmonic of radio emission.
  \item  The radio flux of the burst attained $F=  10^3$ s.f.u.
  \item  The unusual burst was observed at a frequency as low as 1 MHz (STEREO/WAVES data).
\end{enumerate}

We propose the following scenario of the generation of the caterpillar burst.
The unusual burst was associated with
active region NOAA1222 which started to be active at about 11:36
UT. The magnetic flux rope, which believed to be associated with this
activity,  propagated radially with a velocity of about 500 km s$^{-1}$
away from the Sun. At heights $R=2.22 - 2.55 \: R_\mathrm{s}$ the magnetic
flux rope accelerated electrons. Moving towards and
away from the Sun these particles generated the radio emission
which was observed as the unusual radio burst. The particles
propagating towards the Sun produced the radio emission with drift
from low to high frequencies whereas particles moving away from
the Sun generated the radio emission drifting from high to low
frequencies. The velocities of the accelerated particles changed
from  $0.8 \cdot 10^9$ cm s$^{-1}$ to $1.4 \cdot 10^9$ cm s$^{-1}$ . At
frequencies $>27.5$ MHz the unusual burst was not observed because
its radio emission could not propagate through the plasma of the
solar corona towards the Earth.

\begin{acks}
The work was partially fulfilled in the frame of FP7 project
SOLSPANET (FP7-PEOPLE-2010-IRSES-269299).  M. Panchenko
acknowledges the Austrian Fond zur Foerderung der
wissenschaftlichen Forschung (FWF projects P23762-N16).
\end{acks}

\appendix
\section{Propagation of the Radio Emission at the Second Harmonic from Behind-the-limb Region}

Here we discuss the possibility for the radio emission at the second harmonic
of local plasma frequency to propagate from the behind-the-limb
region of the Sun towards the Earth. As was discussed in the present paper
the unusual burst seems to be generated at the second harmonic of local plasma
frequency. This follows from low degree of polarization of the
unusual radio burst which was about 10\%. Note that the fundamental radio emission
had higher degree of polarization (type IIIb bursts have 60-80\%).

Let us discuss in detail the possibility to detect the radio emission at the second
harmonic propagating from a behind-the-limb region by a ground-based radio telescope.
Figure 13 presents the scheme of such propagation.  The radio source is situated
at a distance $R_\mathrm{H}$  from the solar center with the longitude $\alpha$. The
second harmonic can propagate towards the Earth if the distance
$R_\mathrm{l}$  is larger than the distance $R_\mathrm{F}$. The last quantity corresponds to
the radial distance at which the local plasma frequency
$\omega_\mathrm{F}$  in the solar corona equals the double plasma frequency
$\omega_\mathrm{H}= 2 \omega_\mathrm{pe}$  at the place of generation of the
second harmonic $\omega_\mathrm{F}  \: (R =\: R_\mathrm{F})  =  \omega_\mathrm{H}  \: (R=
\: R_\mathrm{H}) =2 \omega_\mathrm{pe} $. Therefore the condition for the
observation of the behind-the-limb radio emission is
 \begin{equation}\label{A1}
   R_\mathrm{l} > R_\mathrm{F}.
 \end{equation}
In the opposite case, when $R_\mathrm{l} < R_\mathrm{F}$ , the radio emission
at the second harmonic cannot propagate towards the Earth.

In the case of the Newkirk model \cite{Newkirk1961} of the solar corona,
$ n =n_\mathrm{N} \cdot 10^{4.32/r}$, where $n_\mathrm{N}= 4.2 \cdot 10^4$ cm$^{-3}$ and $r=R/R_\mathrm{s}$,
Equation (\ref{A1}) is satisfied when

 \begin{equation}\label{A2}
    R_\mathrm{H}> \frac{2.16}{\log{2}} \left( \frac{1}{\cos{\alpha}} - 1  \right)  R_\mathrm{s}.
 \end{equation}
For the Baumbach-Allen model \cite{Allen1947}, $n=n_\mathrm{BA} (1.55 r^{-6} + 2.99 r^{-16})$,
where $ n_\mathrm{BA} = 10^8 \: \mathrm{cm}^{-3}, \: r=R/R_\mathrm{s} $, we have

 \begin{equation}\label{A3}
    \cos{\alpha}> 2 ^{-1/3}
 \end{equation}
(for simplicity in the case of distances $r>1.5$  we take into account only
the first term, \emph{i.e.} $n=n_\mathrm{BA} \cdot 1.55 r^{-6}$). A similar result is obtained
for the Leblanc model \cite{Leblanc1998}.

Therefore we can define the regions in the ``frequency -
longitudinal angle" plane where the radio emission at the second
harmonic can propagate toward the Earth. In the case of the
Baumbach-Allen  model the emission can be detected at the Earth if
the source is at the longitudinal angles $\alpha < 37.5^ {\circ}$.
Moreover the radio emission can be observed in the whole frequency
range (Figure 14, to the left from the dashed line).

If the coronal density follows the Newkirk model the higher
possible frequency (or higher cut-off frequency) of the radio
emission at the second harmonic detected at the Earth depends on the
longitudinal angle $\alpha$ (Figure 14, solid line). Therefore in
this case the radio emission can be observed in regions to the left of the
solid line in Figure 14. This line is defined by

 \begin{equation}\label{A4}
    f^* = \frac{1} {\pi} \left( \frac{4 \pi e^2 n_\mathrm{N}} {m_\mathrm{e}} \right) ^{1/2} 2 ^{\cos{\alpha}/(1 - \cos{\alpha})}
 \end{equation}
where ${e}$ and ${m_\mathrm{e}}$ is the charge and mass of the electron.

Assuming that the source of the unusual radio burst was related to
active region NOAA1222, which was located at the longitudinal
angles $40^{\circ} -50^{\circ}$, and using Equation (\ref{A2}) we
derive the distances at which the second harmonic emission
can come out. These distances are  $2.16 \cdot
R_\mathrm{s}$ (for $\alpha = 40^{\circ}$) and $3 \cdot R_\mathrm{s}$ (for $\alpha =
50^{\circ}$). Maximal frequencies of the radio emission that
are able to come out from these heights are 37 MHz and 12.7 MHz
respectively. The cut-off frequency $f^*=27.5$ MHz corresponds
to the height $2.48 \cdot R_\mathrm{s}$ is the highest
frequency of the observed unusual radio burst.

%%% BIBLIOGRAPHY %%%%%%%%%%%%%%%%%%%%%%%%%%%%%%%%%%%%%%%%%%%%%%%%%%%%%%%%%%%

\bibliographystyle{spr-mp-sola}
%\bibliography{Unusual_bursts_Melnik}

%\begin{thebibliography}{}
% \bibitem[\protect\citeauthoryear{{Berger}}{2003}]{Berger03b}
%Berger,~M.A.:2003, in Ferriz-Mas, A., N{\'u}{\~n}ez, M. (eds.),
%    \textit{Advances in Nonlinear Dynamics}, Taylor and Francis Group,
%    London, 345.

%\end{thebibliography}

\newpage

 %%%%%%%%%%%%%%%%%%%%%%%%%%%%% FIGURES   %%%%%%%%%%%%%%%%%%%%%%%%%%%%%%%%%%%%
 \begin{figure}    %%%%%%%%%%%%%%%%%% FIGURE 1
   \centerline{\includegraphics[width=0.7\textwidth,clip=]{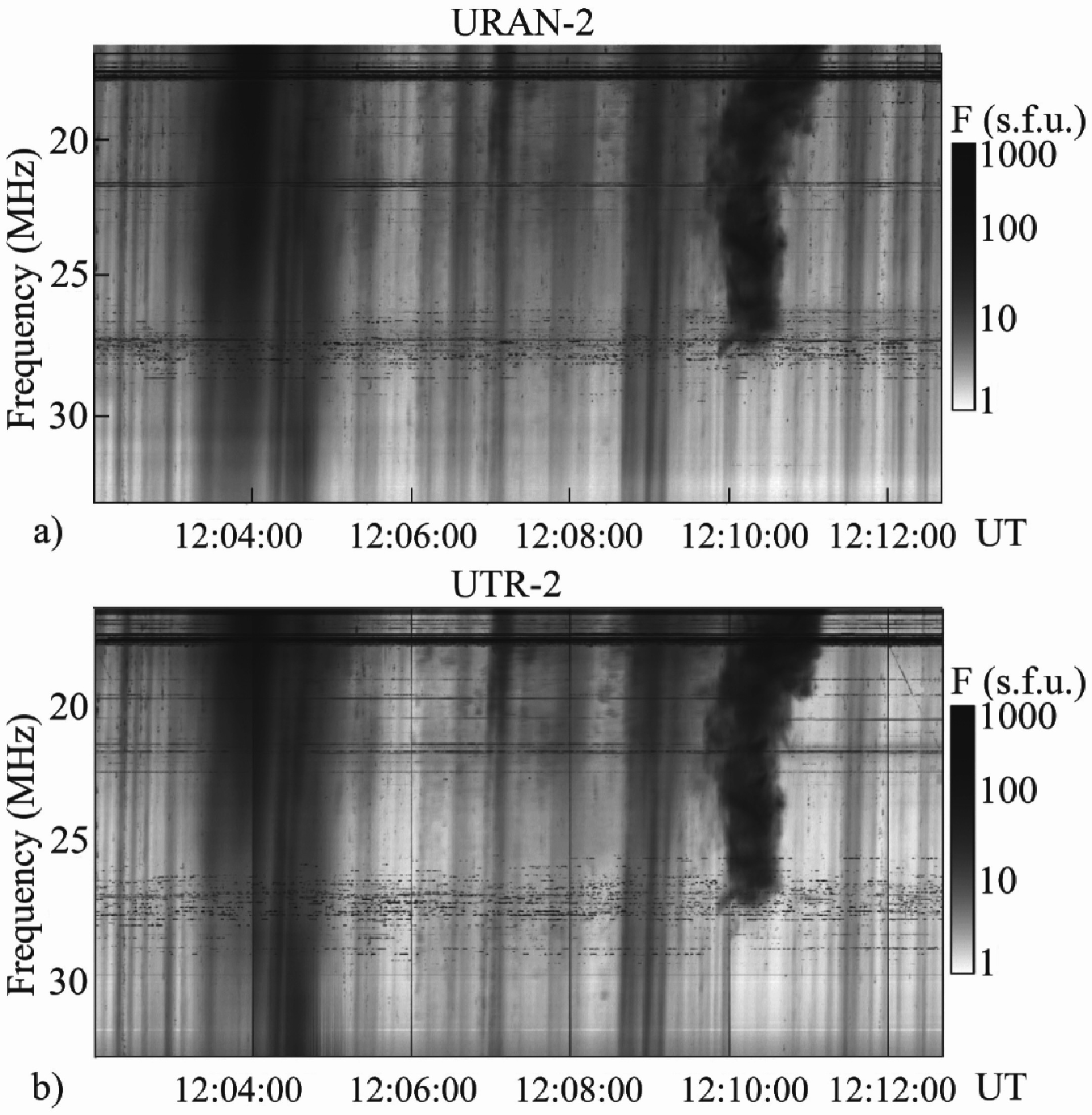}}
      \caption{Unusual burst in the form of a caterpillar (12:10:00) observed by URAN-2 (a)
      and UTR-2 (b) against the background of a type III storm on 3 June 2011.}
   \label{fig1}
   \end{figure}

  \begin{figure}    %%%%%%%%%%%%%%%%%% FIGURE 2
   \centerline{\includegraphics[width=0.7\textwidth,clip=]{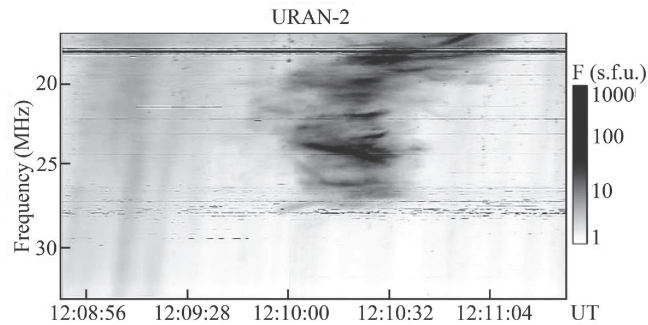}}
      \caption{Fine structure in a form of horizontal stripes is well visible in the
               high resolution radio spectra of the unusual burst observed by URAN-2.}
   \label{fig2}
   \end{figure}

   \begin{figure}    %%%%%%%%%%%%%%%%%% FIGURE 3
   \centerline{\includegraphics[width=0.8\textwidth,clip=]{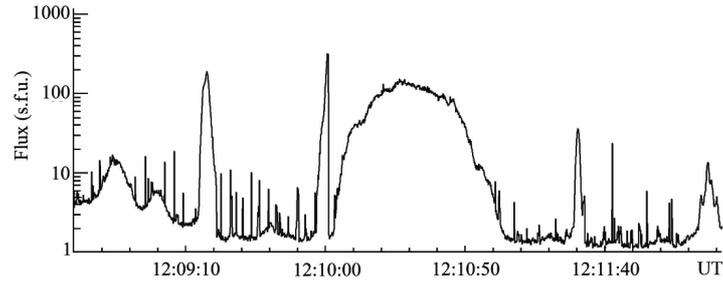}}
      \caption{Time profile of the unusual burst at 26 MHz.}
   \label{fig3}
   \end{figure}

   \begin{figure}    %%%%%%%%%%%%%%%%%% FIGURE 4
   \centerline{\includegraphics[width=0.8\textwidth,clip=]{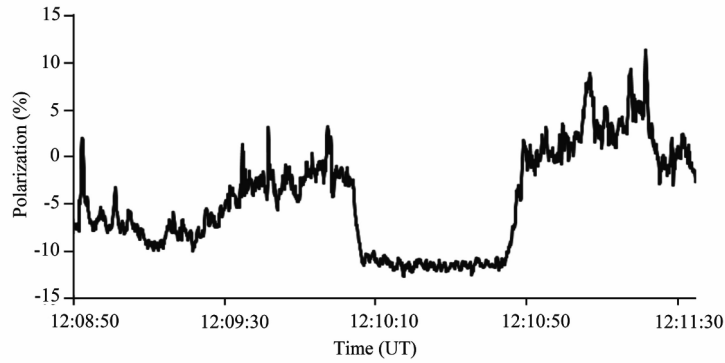}}
      \caption{Polarization of the unusual burst at 26 MHz according to URAN-2 observations.}
   \label{fig4}
   \end{figure}

   \begin{figure}    %%%%%%%%%%%%%%%%%% FIGURE 5
   \centerline{\includegraphics[width=1\textwidth,clip=]{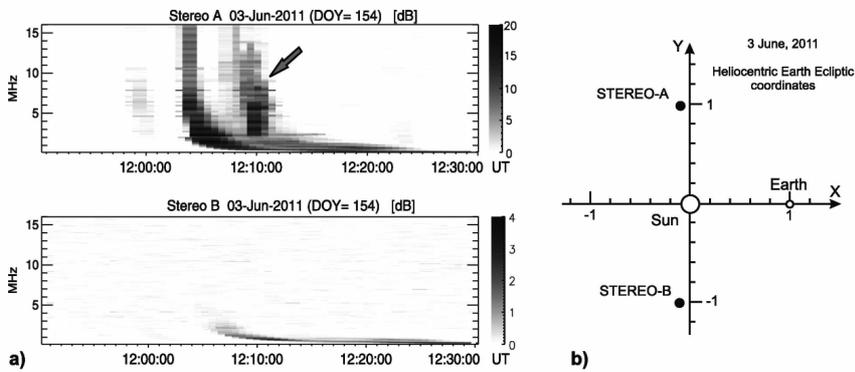}}
      \caption{Radio spectra obtained by WAVES experiment onboard STEREO-A and STEREO-B on 3 June 2011 (panel (a)).
                The unusual solar radio burst  was observed only by STEREO-A (marked by arrow). Panel (b) shows the
                positions of spacecraft in the heliocentric earth ecliptic coordinates.}
   \label{fig5}
   \end{figure}

   \begin{figure}    %%%%%%%%%%%%%%%%%% FIGURE 6
   \centerline{\includegraphics[width=0.8\textwidth,clip=]{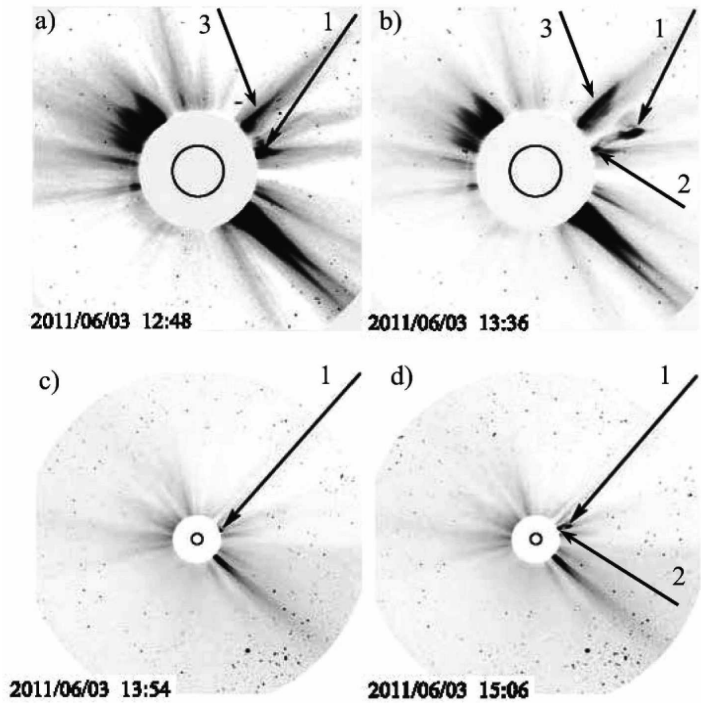}}
      \caption{Two ejections (1 and 2) and jet (3) observed by LASCO C2 ((a), (b)) and by LASCO C3 (c), (d)) at the consistent moments of time.}
   \label{fig6}
   \end{figure}

   \begin{figure}    %%%%%%%%%%%%%%%%%% FIGURE 7
   \centerline{\includegraphics[width=1\textwidth,clip=]{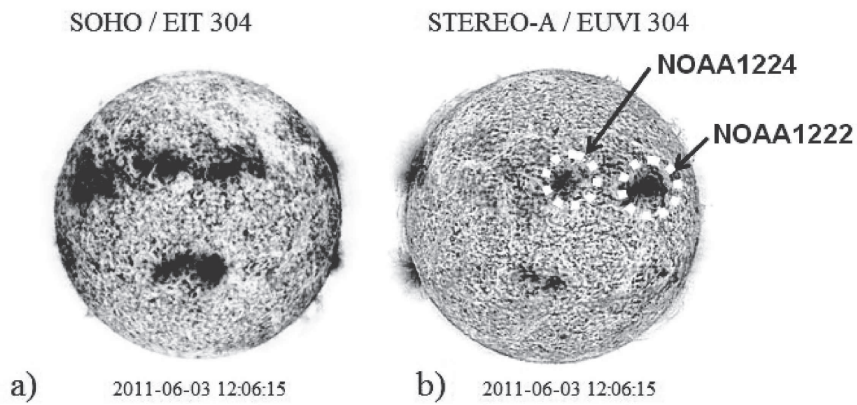}}
      \caption{The solar disk with active regions observed by SOHO Extreme ultraviolet Imaging Telescope (SOHO/EIT 304{\AA}) (a) and STEREO-A Extreme UltraViolet Imager (SECCHI EUVI) (b).}
   \label{fig7}
   \end{figure}

    \begin{figure}    %%%%%%%%%%%%%%%%%% FIGURE 8
    \centerline{\includegraphics[width=0.8\textwidth,clip=]{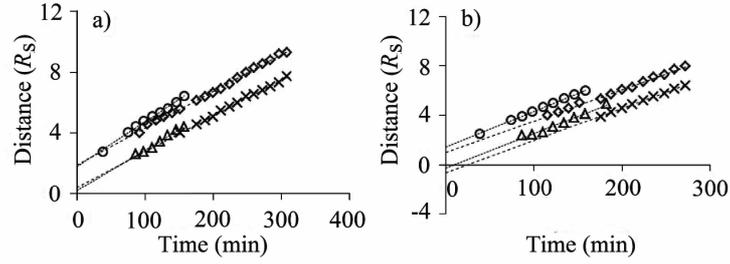}}
      \caption{Motions of the front (a) and center (b) of ejections 1 (circles and rhombuses) and 2 (triangles and crosses).
      The time 0 in Figures 8-10 corresponds to 12:10 UT.}
   \label{fig8}
   \end{figure}

     \begin{figure}    %%%%%%%%%%%%%%%%%% FIGURE 9
   \centerline{\includegraphics[width=0.6\textwidth,clip=]{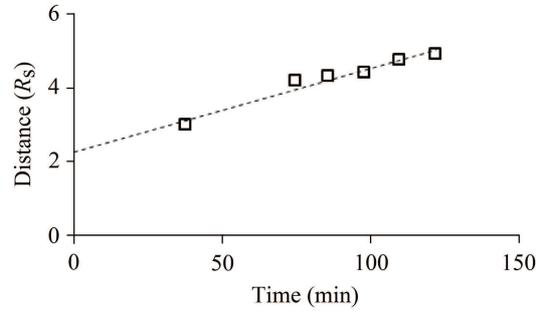}}
      \caption{The motion of the jet front in the sky plane according to LASCO C2.}
   \label{fig9}
   \end{figure}

        \begin{figure}    %%%%%%%%%%%%%%%%%% FIGURE 10
   \centerline{\includegraphics[width=0.6\textwidth,clip=]{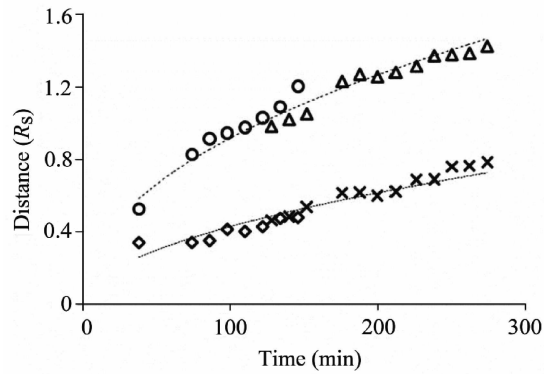}}
      \caption{Longitudinal (circles and triangles) and transverse (rhombuses and crosses) sizes of the first ejection according to LASCO C2 and C3 data.}
   \label{fig10}
   \end{figure}

   \begin{figure}    %%%%%%%%%%%%%%%%%% FIGURE 11
   \centerline{\includegraphics[width=0.6\textwidth,clip=]{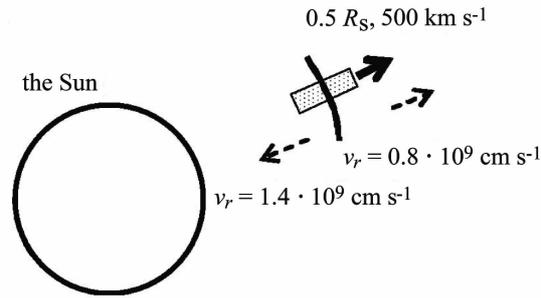}}
      \caption{Scheme of the propagation of a magnetic flux rope in the solar corona.}
   \label{fig11}
   \end{figure}

  \begin{figure}    %%%%%%%%%%%%%%%%%% FIGURE 12
   \centerline{\includegraphics[width=1\textwidth,clip=]{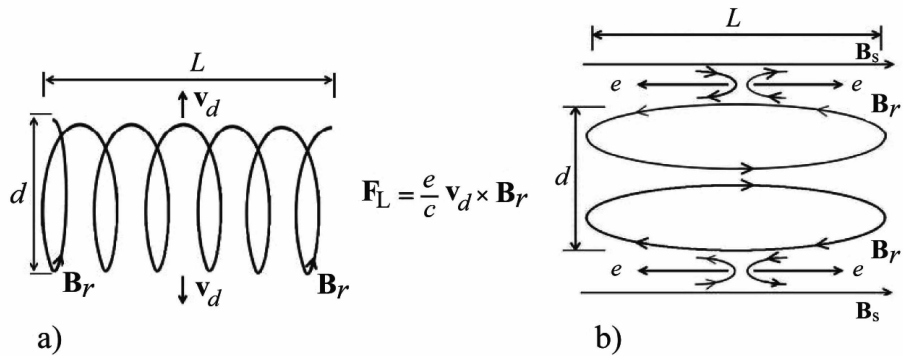}}
      \caption{Acceleration of electrons by magnetic bunch due to the Lorenz force (a) and reconnection processes (b).}
   \label{fig12}
   \end{figure}

   \begin{figure}    %%%%%%%%%%%%%%%%%% FIGURE 13 (Appendix Fig1)
   \centerline{\includegraphics[width=0.4\textwidth,clip=]{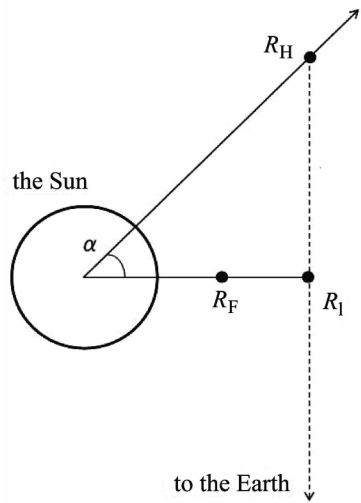}}
      \caption{Scheme of propagation of radio emission at the second harmonic $\omega_\mathrm{H} = 2 \omega_\mathrm{pe}$ (at altitude $R_\mathrm{H}$)  in the direction ``towards the Earth"  (latitudinal plane).}
   \label{fig13}
   \end{figure}

   \begin{figure}    %%%%%%%%%%%%%%%%%% FIGURE 14 (Appendix Fig2)
   \centerline{\includegraphics[width=0.6\textwidth,clip=]{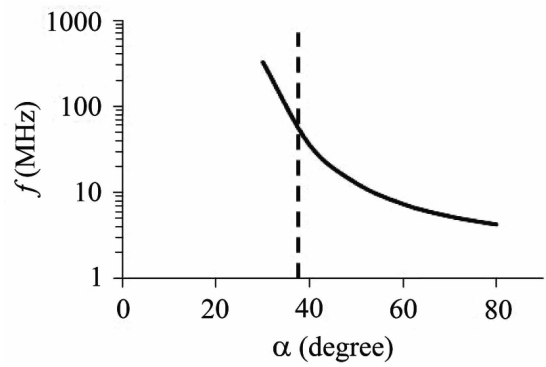}}
      \caption{Regions of propagation  of the second harmonic radio emission
                to the Earth from the ``behind-the-limb" hemisphere in the Baumbach-Allen
                (dashed line) and Newkirk (solid line) models.}
   \label{fig14}
   \end{figure}
--------------------------------------------------

\end{article}

\end{document}